\documentstyle{aipproc}
\input boxedeps.tex
\SetRokickiEPSFSpecial  
\HideDisplacementBoxes

\newcommand{\D}{\overline{D}}
\newcommand{\da}{\dagger}  
\newcommand{\be}{\begin{equation}}
\newcommand{\eq}{\end{equation}}
\newcommand{\Tr}{{\rm \, Tr \!}}    
\newcommand{\dm}{{\cal M}}          
\newcommand{\newl}{l}               

\begin{document}
\mbox{}\hfill DAMTP-99/171\\
\title{ Introduction to \\
Transverse Lattice Gauge Theory\footnote{Invited lectures at
APCTP International Light-Cone School {\em New Directions in
Quantum Chromodynamics}, May 26 - June 26 1999, Seoul, Korea.}}

\author{Simon Dalley}
\address{Department of Applied Mathematics and Theoretical Physics \\
Silver Street, Cambridge CB3 9EW, England}

\maketitle

\begin{abstract}
I review a new treatment of an old idea for
light-front quantization of lattice gauge theories and give 
new results from some illustrative calculations:
[I] Transverse Lattice Gauge Theory;
[II] Pure Glue;
[III] Heavy Sources and Winding Modes;
[IV] An Example -- Large-$N$ QCD in $2+1$ Dimensions.
\end{abstract}

\section{Transverse Lattice Gauge Theory}

\subsection{Light-Front Co-ordinates}

In the constituent quark model, hadrons are composed of a few 
quarks moving about in empty space and bound by non-relativistic
potentials. It is an {\em extremely} successful model.
Before 
the advent of QCD  as the quantum field theory of hadron structure,
there was little reason to think of hadrons in any other way. But while 
there is consensus that QCD is fundamentally correct, the low-energy dynamics 
of this gauge theory are thought to involve a complicated vacuum with
arbitrary numbers of relativistic quarks and gluons. How could these
pictures be reconciled? Indeed, how can one efficiently tackle the
problem of relativistic
strongly bound states at all?

There is a popular yet poorly understood
formulation of quantum field theory --- Light-Front
quantisation  --- where something like the
constituent picture of boundstates arises naturally.
In fact, it offers
much more than that, since it 
furnishes a Schrodinger equation suitable for relativistic many-body
systems.

In the co-ordinate system $\{ x^0, x^1, x^2, x^3 \}$, where
$x^0$ is time (I set $c = \hbar = 1)$, 
one usually initializes the wavefunction $\Psi(x^1,x^2,x^3)$ on a 
hyper-plane $x^0 = {\rm constant}$. 
If the four-momentum of the system is 
$\{P^0, P^1, P^2, P^3 \}$, the energy-operator $\hat{P}^0$ 
is the equal-$x^0$ hamiltonian that  evolves the wavefunction 
in $x^0$ according to Shrodinger's equation 
\be
{\rm i}  {\partial \Psi  \over \partial x^0} = \hat{P}^0 \Psi.
\eq
This formulation has manifest rotational invariance and is tractable for most
non-relativistic problems in physics. 
It is not so suitable when the momenta $P^{i=1,2,3} \sim P^0$ (recall that
$P^0$ contains the rest-energy). 

In light-front (LF) quantisation \cite{dirac} on the other 
hand, we (conventionally) define
\be
x^{\pm} = {x^0 \pm x^3 \over \sqrt{2}} \ , \ 
{P}^{\pm}= {{P}^0 \pm {P^3} \over \sqrt{2}} \ . \label{new}
\eq
$P^{\pm}$ is canonically conjugate to $x^{\mp}$, and we interpret $(x^-, P^+)$
as space and momentum, while $(x^+, P^-)$ as time and energy. ${\bf x}
= \{x^1, x^2 \}$ and ${\bf P} = \{ P^1 , P^2 \}$ are the transverse
co-ordinates.
The Minkowski metric in the new co-ordinate system is such that the
invariant length
\be
ds^2 = 2 dx^+ dx^- - (dx^1)^2 - (dx^2)^2 \ 
\eq
and $x^+ \equiv x_-$.
The  wavefunction $\Psi_{\rm lc}$ is initialized on a null hyper-plane 
$x^+ = {\rm constant}$ and
the LF Schrodinger equation that evolves
$\Psi_{\rm lc}(x^-,x^1,x^2)$ 
in LF time
$x^+$ is
\be
{\rm i}  {\partial \Psi_{\rm lc}  \over \partial x^+} = \hat{P}^-
\Psi_{\rm lc}.
\eq
This equation is now  suitable
for relativistic problems. For example, $\Psi_{\rm lc}$ is
manifestly Lorentz-boost invariant;  a detailed discussion of the advantages
and historical survey of LF ideas can be found in ref.\cite{review}.

It is  easy to see why LF co-ordinates may help with the 
constituent question, 
if we consider
the energy-momentum relation for a (free) particle of mass $\mu$ 
\be
P^- = {(P^1)^2 + (P^2)^2 + \mu^2 \over 2 P^+} \ . \label{disp}
\eq
With the conventional interpretation
of positive energy ($P^0 \geq 0$) for  particles and anti-particles, 
it follows that light-front momentum $P^+ \geq 0$.
Because $P^+$ is conserved and the total $P^+$ of the vacuum
must be zero (by translation
invariance), 
each particle contributing to the vacuum state 
can have only $P^+ = 0$. But according to
to (\ref{disp}), these `zero modes' have infinite light-front energy. If
we introduce a high energy cut-off, as is always necessary for
defining a quantum field theory, 
these zero modes and the vacuum structure they carry will
be explicitly removed. Their effects on observables
below the cut-off should be felt through renormalisation of the
hamiltonian. If this
renormalisation entails quarks getting a constituent mass, and the
gluon degrees of freedom getting at least a mass gap, if not constituent
masses themselves, then we have arrived at the basic picture of the
constituent quark model. 

The trouble with the above idea is that the 
LF renormalisation procedure is still quite poorly understood for
theories like QCD, though interesting efforts are being made in this
direction \cite{rg1,rg2,rg3,rg4}. A formulation of quantum field theory where 
non-perturbative effects normally associated with the vacuum have to
appear via explicit counter-terms in the hamiltonian is also alien
to many people. For these reasons
there have been some reservations about the consistency
of 
light-front quantisation. Nevertheless, the ideas are physically
appealing and a number of examples are known where 
one can  derive the renormalisation resulting
from the removal of certain non-dynamical zero modes; 
see Burkardt's lectures at this school in a previous year \cite{mat}.
The attitude I will take in these lectures is that the potential 
applications  of light-front
quantisation are  far too important for it to be simply dismissed.
I will describe a practical framework for doing calculations in LF
QCD with a high-energy cut-off, pioneered by Bardeen and Pearson
\cite{bard1,bard2}, which
uses symmetry to guide renormalisation of the theory. 
The symmetries that define QCD are gauge, Lorentz, and 
chiral invariance. In these lectures, I will only discuss pure gauge
theories and infinitely heavy quarks, so chiral invariance will
not be treated.

The method will initially allow
all operators in the LF hamiltonian that respect
symmetries unviolated by the cut-off --- allowed operators. 
After
systematically truncating to a subset of these allowed operators, one
then tunes the remaining
couplings {\em a posteriori} 
to restore the violated symmetries in observables as best one can. 
This is a physically motivated, systematically improvable
framework which  does not use any fits to experimental data, i.e. it
uses only first principles.  
At present, the tests for violations of Lorentz covariance are done
in a rather inefficient way, simply by
solving a whole bunch of
hamltonians to find the correct one. 
Nevertheless, the recent results obtained in this way \cite{us3,us4,us5} are 
surprisingly accurate. Once
a more efficient treatment of Lorentz covariance is developed, or 
non-perturbative LF hamiltonian renormalisation group ideas are better
understood, many new areas of non-perturbative QCD will be
opened up.

\subsection{Transverse Lattices}

Light-front quantisation already has more manifest Lorentz symmetry than any
other hamiltonian quantisation scheme \cite{dirac}, but typically one cannot
avoid breaking rotational invariance; the choice $x^3$ in (\ref{new}), rather
than $x^1$ or $x^2$, is arbitrary. On the other hand, we would like
to preserve gauge invariance (later this will also greatly facilitate
the treatment of confinement).
This can be done with a lattice 
cut-off \cite{wilson,suss}. In a hamiltonian formulation time remains continous
and infinite, of course. 
For a LF hamiltonian $\hat{P}^-$, whose configuration space
is $\{x^-, x^1, x^2\}$, only lattice discretization of the
transverse directions ${\bf x}$ is
appropriate.
Discretizing the longitudinal co-ordinate $x^-$ is not
appropriate since it would cut-off large values of the
conjugate variable $P^+$; but it's small
values of $P^+$ that correspond to high energy. To remove the
high-energy region in a gauge-invariant way
we can impose, say,  anti-periodic 
boundary conditions on  $x^-$ \cite{casher}.
In the following we use indices, $\mu,\nu \in
\{0,1,2,3,\}$, $\alpha, \beta \in \{+,-\}$ and $r \in \{1,2\}$,
and sum over repeated indices. We introduce a transverse lattice spacing $a$ 
and a longitudinal period ${\cal L}$.

To construct an $SU(N)$ transverse lattice (pure) gauge theory on this
spacetime, we introduce gauge potentials $A_{\alpha}$ in the Lie algebra
of $SU(N)$ for the continuum directions, and lattice variables
for the transverse directions. The particular choice of lattice
variables which will be convenient for LF quantisation
are complex $N$x$N$ matrices $M_{r}(x^+,x^-,{\bf x})$
associated with the transverse
link from ${\bf x}$ to ${\bf x} + a \hat{\bf r}$, at position $(x^+,
x^-)$ --- colour-dielectric link variables. Because they are linear 
variables, it is easy to identify the independent degrees of freedom. 
It is sometimes helpful to think of them
as an average over paths ${\cal C}$ between these points, with some
weight $\rho ({\cal C})$, of the short-distance continuum gauge potentials
\be
M_r = \sum_{\cal C} \rho ({\cal C}) {\rm P}\  {\rm exp}\left\{ {\rm i}
\int_{\bf x}^{{\bf x} +a \hat{\bf r}}
A_{\mu} dx^{\mu} \right\}
\eq
Near the continuum limit $a=0$, the potentials must change
only very slowly over many lattice spacings, and $M$ must be forced
to lie in the $SU(N)$ group. At larger $a$ there is no such restriction
however. In fact, for non-abelian gauge theories, it makes more
physical 
sense to use disordered complex variables $M$ when the lattice cut-off
$a$ is quite large. 

Continuum gauge transformations induce the following lattice gauge
transformations on these variables
\begin{eqnarray}
        A_{\alpha}({\bf x}) & \to & V({\bf x}) A_{\alpha}({\bf x}) 
        V^{\da}({\bf x}) + {\rm i} \left(\partial_{\alpha} V({\bf x})\right) 
        V^{\da}({\bf x}) \ , \\
        M_r({\bf x}) &  \to & V({\bf x}) M_r({\bf x})  
        V^{\da}({\bf x} + a\hat{r})   \; ,
\end{eqnarray}
where $V \in SU(N)$. The strategy will be to construct LF
hamiltonians invariant under these gauge transformations and 
any Lorentz symmetries that have not be violated by the cut-offs;
this include boosts along $x^3$ and discrete $Z_4$ rotations about $x^3$.
We will seek to enhance the remaining Lorentz symmetries, 
generically broken by the cut-offs,
by tuning the couplings of these hamiltonians.

The continuum QCD action is
\be
- \int dx^4 {1 \over 2 g^2} \Tr\left\{ F_{\mu \nu}F^{\mu \nu}\right\} \
. \label{cont}
\eq
The following gauge-invariant transverse lattice 
action is the simplest that reduces to (\ref{cont}) in the naive
continuum limit $a \to 0$, ${\cal L} \to \infty$, \cite{bard2}
\begin{eqnarray}
S &  = & \int dx^+ dx^- \sum_{{\bf x}} 
  \Tr\left\{ \D_{\alpha} M_r({\bf x}) \left(\D^{\alpha} 
M_r({\bf x})\right)^{\da}
\right\} \nonumber \\ 
&& -  {a^2 \over 2g^2} \Tr\left\{F_{\alpha\beta}F^{\alpha\beta}\right\}
+ {\beta \over N a^2}  \Tr \left\{ M_{\rm plaq} + 
M_{\rm   plaq}^{\dagger} 
\right\} - U[M] 
\label{lag}
\end{eqnarray}
where
\be
        \D_{\alpha} M_r({\bf x})  
        =  \left(\partial_{\alpha} +i A_{\alpha} ({\bf x})\right)
        M_r({\bf x})  
        -  i M_r({\bf x})   A_{\alpha}({{\bf x}+a\hat{\bf r}}) \;,
\label{covdiv}
\eq
$M_{\rm plaq}$ is the product of link matrices around a transverse
plaquette,
and $U[M]$ is any gauge-invariant 
potential that forces $M$ into $SU(N)$ as $a \to 0$,
such as
\be
U[M] = 
{N  \over \lambda} \left( \Tr \left\{
(1 - M^{\da}_r ( {\bf x})  M_r ({\bf x}))^2 \right\} 
+ (\det{M} - 1)^2 \right)
\ ,
\eq 
where $\lambda \to 0$ as $a \to 0$.

More generally, $S[M]$ can 
consist of all the combinations of $M, F^{\alpha \beta}, 
\bar{D}^{\alpha} M$ which are invariant under gauge and 
residual Lorentz symmetries. 
Their couplings are to be chosen so 
as to restore the symmetries violated by the cut-offs.

\section{Pure Glue}

\subsection{Colour-Dielectric Expansion}

$S[M]$ contains an infinite number of allowed operators, so to begin
calculations we need to truncate them to a finite set in some
physically reasonable way.
Of the two cut-offs, $a$ and ${\cal L}$, it will be possible to 
extrapolate the latter. It makes sense therefore to restrict
allowed operators on
dimensional grounds w.r.t. $\{0,3\}$ co-ordinates. The 
divergences that appear as ${\cal L} \to \infty$ are
of normal-ordering type, and thus easily dealt with.
We will also demand an interaction-independent (kinetic)
momentum operator $P^+$ and naive LF parity $x^+ \leftrightarrow x^-$.
The main further restriction will be to expand the LF
hamiltonian $\hat{P}^-$, that results from $S[M]$, in powers of
$M$ --- the `colour-dielectric expansion'. 
This only makes sense if, for a given $a$,
 the $P^-$ that best recovers Lorentz covariance 
is analytic about $M=0$ and (classically) minimized there.
(This could not be the case in a neighborhood of $a=0$, where $M$
should
be forced into $SU(N)$). We 
are not guaranteed to find anything at all. There is,
however, circumstantial evidence from Euclidean lattice work
\cite{pirner} that a Lorentz covariant
scaling trajectory in the space of couplings flows from
the continuum limit to a large $a$ region where $M=0$ is
the classical minimum. Our job is to explictly verify
the existence of such a trajectory for the transverse lattice by testing this
region for signs of Lorentz covariance restoration.

\subsection{Gauge-Fixing}

Light-front quantization is greatly simpled by the LF gauge choice
$A_{-} = 0$.\footnote{Because we use anti-periodic $x^-$ boundaries,
the $\int dx^- A_{-} $ zero mode has been removed from the theory.
Its omission can be shown to renormalise the mass of the $M$ field.}
This still leaves $x^-$-independent gauge transformations unfixed.
Suppose we start with an action
\begin{eqnarray}
S &  = & \int dx^0 dx^3 \sum_{{\bf x}}  \left(
  \Tr\left\{ \D_{\alpha} M_r({\bf x}) \left(\D^{\alpha} 
M_r({\bf x})\right)^{\da}
\right\} \right. \nonumber \\
&& \left. -  {1 \over 2G^2} \Tr\left\{F_{\alpha\beta}F^{\alpha\beta}\right\}  
- V_{{\bf x}}[M]\right) \;,
\label{lag}
\end{eqnarray}
where $V[M]$ contains all gauge-invariant products of $M$ up
$O(M^4)$ and $G^2 \to g^2 / a^2$ as $a \to 0$. 
This is still not yet a finite number of operators, but we can obtain a
finite
set by further requiring a degree of {\em transverse locality} for
operators
that are products of gauge-invariant operators (see later).

This action in LF gauge reduces to
\begin{eqnarray}
S(A_{-} = 0) &  = & \int dx^0 dx^3 \sum_{{\bf x}}  \left(
  \Tr\left\{ \partial_{+} M_r({\bf x}) \partial_{-} 
M_r({\bf x})^{\da}
\right\} + {\rm c.c.}\right. \nonumber \\
&& \left. +  {1 \over G^2} \Tr\left\{ (\partial_{-} A_{+} )^2\right\}
+ \Tr\left\{  A_{+} J^{+}({\bf x}) \right\}  
- V_{{\bf x}}[M]\right) \;,
\label{lag}
\end{eqnarray}
where
\be
 J^{+} =  i \left(
M_r ({\bf x}) \stackrel{\leftrightarrow}{\partial}_{-} 
M_r^{\da}({\bf x})  + M_r^{\da}({\bf x} - a \hat{\bf r})
\stackrel{\leftrightarrow}{\partial}_{-} M_r({\bf x} - a \hat{\bf r})
   \right) \ . \label{jay}
\eq
We note that the equation of motion for $A_{+}$ is a constraint (no
$x^+$-derivatives)
\be
{2 \over G^2} (\partial_{-})^{2} A_{+} = J^+ - {1 \over N} \Tr \ J^+ \ ,
\eq
which can be inverted for $A_{+}$
\be
A_{+} = {G^2 \over 2} {1 \over \partial_{-}^{2}} \left( J^+ - 
{1 \over N} \Tr \ J^+ 
\right) \ .
\eq

\subsection{Canonical LF Quantisation}
The following canonical momenta can be
straightforwardly 
derived in gauge-invariant form
 from the usual definition in terms of the energy momentum
tensor $P^{\alpha}= \int dx^- \sum_{{\bf x}} T^{+ \alpha}$ ;
\begin{eqnarray}
P^+ & = & 2 \int dx^- \sum_{{\bf x}}  \Tr\left\{ \partial_{-} M_r({\bf
x}) \partial_{-} M_r({\bf x})^{\da} \right\}  \\
P^-  & = & \int dx^- \sum_{{\bf x}}
 V[M] - {G^2 \over 4} \Tr\left\{  J^{+} \frac{1}{\partial_{-}^{2}}
J^{+} \right\} + {G^2 \over 4N} \Tr \ J^{+} \frac{1}{\partial_{-}^{2}}
\Tr \ J^{+} \ .\label{ham}
\end{eqnarray}
The transverse momenta ${\bf P}$ and boost-rotations $M^{\mu \nu}$ will be
discussed shortly (see ref.\cite{review} for a summary of important
properties of the Poincar\'e
generators in light-front formalism).
$M_r$ is conjugate to $\partial_{-} M^{\da}_{r}$, and we impose
equal-$x^+$
commutation relations
\be
        \left[M_{r,ij}(x^-,{\bf x}), 
        \left(\partial_- M_{s,kl}(y^-,{\bf y})\right)^\da\right]
        = {1 \over 2} \delta_{il}\,\delta_{jk}\, \delta (x^- -y^-)
        \,{\bf \delta_{x, y}} \,\delta_{r,s} \;.
\eq
where $i,j \in \{1, \cdots N \}$ are colour indices.

It is useful to make a Fourier decomposition at $x^+ = 0$
with respect to $x^-$ in
terms of free fields
\be
 M_r(x^+=0,x^-,{\bf x})   =  
        \frac{1}{\sqrt{4 \pi }} \int_{0}^{\infty} {dk^+ \over \sqrt{ k^+}}
        \left( a_{-r}(k^+,{\bf x})\, e^{ -i k^+ x^-}  +  
        a^{\da}_r(k^+,{\bf x})\, e^{ i k^+ x^-} \right) 
\eq
that induces
\begin{eqnarray}
   \left[a_{\lambda,ij}(k^+,{\bf x}), 
        a_{\rho,kl}^{*}(\tilde{k}^+, {\bf y})\right] 
        & = & \delta_{ik}\, \delta_{jl}\, \delta_{\lambda \rho}\, 
        {\bf \delta_{x, y}}\,\delta(k^+-\tilde{k}^+) \;, \\
   \left[a_{\lambda,ij}(k^+,{\bf x}),
        a_{\rho,kl}(\tilde{k}^+, {\bf y})\right] & = & 0 \; ,
\end{eqnarray}
$\lambda, \rho \in \{ \pm 1, \pm 2\}$,  
$a_{\lambda,ij}^{*} = (a^{\da}_{\lambda})_{ji}$.
Acting on a Fock vacuum $|0>$ defined by
\be
a_{\lambda,ij}(k^+, {\bf x} )|0> = 0 \ \ \ \ \forall 
\ k^+ ,{\bf x}, i,j,\lambda \ ,
\eq 
$a_{\lambda,ij}^{*}(k^+,{\bf x}) $ creates a link-parton on $({\bf x},
{\bf x}+a\hat{\bf \lambda})$
on the transverse lattice and carrying
longitudinal momentum $k^+$. (This mixed momentum-coordinate
representation will be 
useful for displaying confinement shortly.)
The general Fock space then splits into blocks of definite total $P^+$
\be
\{ a^* (P^+) |0> , a^* (P^+ - k^+)  a^* (k^+) |0>, \cdots \} \ .
\eq
States of definite ${\bf P}$ will be constructed shorty. The above
Fock basis can be used to write down a matrix representation of
$\hat{P}^-$, which can then be diagonalised to obtain LF
wavefunctions $\Psi_{\rm lc}$. 

\subsection{Physical States}

The first key point to note at this stage is that, because of
positivity $k^+ > 0$, $P^-$ and $P^+$ contain no terms with only
creation operators $a^*$. This means that  the Fock vacuum
satisfies $: P^- : | 0 > = : P^+ : | 0 > = 0$. Of course, this is
only true so long as $M$ is a massive degree of freedom; this is
the region of couplings we will be exploring for Lorentz
covariance. Provided no tachyons appear in the physical spectrum
of boundstates, $|0>$ is the physical vacuum of the theory (in the
presence of our cutoffs). 

Fock space is further simplified by confinement \cite{bard1}. 
Since the operator corresponding to $k^+$ is $\partial_{-}$, a glance
at (\ref{ham}) shows that eliminating $A_{+}$ has introduced
small-$k^+$ singularities. These are cut-off by ${\cal L}$, but will
blow up as ${\cal L} \to \infty$ for fixed $a$ in general. The
divergence
is avoided if the LF charge $Q$ associated to the current $J^+$ (\ref{jay}) 
vanishes
\be
\lim_{k^+ \to 0} \int dx^- {\rm e}^{{\rm i} k^+ x^-} J^+   = 0 \ .
\eq
Only certain Fock states satisfy this condition, $:Q: | {\rm Fock} > = 0$.
They are the states invariant under
the residual
$x^-$-independent gauge transformations
\be
M_r({\bf x})   \to  V({\bf x}) M_r({\bf x})  
        V^{\da}({\bf x} + a{\bf \hat{r}})   \; \ , \ \ \partial_{-} V = 0 \ .
\eq
This will include, for example, closed loops of link-partons on the
transverse lattice 
\begin{eqnarray}
&&\Tr \{ a_{\lambda}^{\da}(P^{+}-k^+,{\bf x}) 
a_{-\lambda}^{\da}(k^+,{\bf x}) \} |0> \ , \nonumber \\
&& \Tr \{ a_{\lambda}^{\da}(P^{+}-k^{+}_{1}-k^{+}_{2}-k^{+}_{3}
,{\bf x}) a_{\lambda}^{\da}(k^{+}_{1},{\bf x} + a \hat{\bf \lambda}) 
a_{-\lambda}^{\da}(k^{+}_{2},{\bf x} + 2a \hat{\bf \lambda}) 
a_{-\lambda}^{\da}(k^{+}_{3},{\bf x} + a \hat{\bf \lambda}) \} |0> \ ,
\nonumber \\
&& {\rm etc..} \nonumber
\end{eqnarray} 
This simple geometrical picture of the finite-energy, gauge-singlets is why
we chose to work in transverse configuration space but longitudinal
momentum space initially.

A remarkable property of the Fock space of connected closed loops of
links\footnote{These are the relevant ones in the $N=\infty$ limit
for example.} is that it is
finite-dimensional in a sector of fixed $P^+$ and ${\bf P}$
for given cut-off ${\cal L}$. If 
we write the momentum of the $i$th parton
as 
\be
k_{i}^{+} = {n_i P^+ \over K} \ ,
\eq
where $n_i \in  \{ 1/2, 3/2, \ldots, K-1/2 \} $ and $K$ is an integer
that plays the role of dimensionless cut-off, the Fock space is
finite-dimensional because individual momenta are discrete and bounded
and the number of partons is bounded (by $2K$). This is the basis
of the numerical treatment known as DLCQ \cite{dlcq}. Matrices of finite
dimension can be calculated and diagonalised, and the results
extrapolated to large $K$.

\subsection{Momentum Eigenstates}

To test Lorentz covariance it will be important to have 
explicit forms for boundstate wavefunctions at non-zero momentum.
So far we have constructed a Fock basis diagonal in $\hat{P}^+$, with
modes localised on transverse links. To obtain states of definite
non-zero transverse momentum ${\bf P}$, we can take a  gauge-singlet 
$p$-link shape, say, and  translate it over
the transverse lattice sites ${\bf y}$ according to
\be
    \sum_{\bf y}
        {\rm e}^{i \bf{P}\cdot \left({\bf y}+{\bf \bar x}\right)} 
      \Tr\left\{ a_{\lambda_1}^\da(k_{1}^{+},{\bf x}_1+{\bf y}) \, 
        a_{\lambda_2}^\da(k_{2}^{+},{\bf x}_2+{\bf y})
        \cdots  a_{\lambda_p}^\da(k_{p}^{+},{\bf x}_p+{\bf y})\right\} 
        \left|0\right\rangle
        \; .  \label{dfunct}
\eq
We must have 
\begin{eqnarray}
\sum_{i=1}^p \widehat{\lambda}_i & = & 0 \nonumber \\
{\bf x}_i & = &  {\bf x}_{i-1} +a \,\widehat{\lambda}_{i-1} \;, 
        \;\;\;\; i<1\le p\; \\
\sum_{i=1}^p k_{i}^{+} & = & P^{+} \nonumber
\end{eqnarray}
and it is convenient to set the phase convention by adopting a
momentum-weighted `centre-of-mass' 
\be
        {\bf \bar x} = \frac{1}{P^+}\sum_{i=1}^p k_{i}^{+} \left({\bf x}_i+
                   \frac{a \, \widehat{\bf \lambda}_i}{2}\right) \ .
               \label{com}
\eq
This is consistent with the canonical gauge-invariant form of the
$M_{-r}$ component of the boost--rotation tensor 
\be
M_{-r} = 2 \int dx^- \sum_{{\bf x}}  (x_r + a \delta_{rs}/2)
\Tr\left\{ \partial_{-} M_s({\bf
x}) \partial_{-} M_s({\bf x})^{\da} \right\}  
\eq
which gives
\begin{eqnarray}
{\rm e}^{-{\rm i}b^r M_{-r}} \, a_\lambda^\da(k^+, {\bf x}) \, 
{\rm e}^{{\rm i}b^r M_{-r}} & = & a_\lambda^\da(k^+, {\bf x})\,
{\rm e}^{-{\rm i}k^+ {\bf b}\cdot ( {\bf x} + a \hat{\bf \lambda}/2)}  \\
{\rm e}^{-{\rm i}b^r M_{-r}}\, |\Psi_{\rm lc}(P^+, {\bf P})\rangle & = & 
|\Psi_{\rm lc}(P^+, {\bf P}- {\bf b} P^+)\rangle \; ,
\end{eqnarray}
Thus,  non-zero ${\bf P}$ states in the Brillouin zone $(-\pi/a, \pi/a)$
can be conveniently obtained from
those at ${\bf P} = 0$.\footnote{The construction of the operator
${\bf P}$ itself is more problematic however,
since the canonical expression is
not gauge-invariant, while the simplest gauge-invariant extensions
are not quadratic.} Longitudinal
boosts generated by 
\be
M_{-+} =  2 \int dx^- \sum_{{\bf x}}  x^- \Tr\left\{ \partial_{-} M_r({\bf
x}) \partial_{-} M_r({\bf x})^{\da} \right\}  
\eq
are even more trivial; up to unimportant global phases 
they simply rescale
longitudinal momentum
\be
{\rm e}^{{\rm i}s M_{-+}}\, |\Psi_{\rm lc}(P^+, {\bf P})\rangle  =  
|\Psi_{\rm lc}({\rm e}^{-s} P^+, {\bf P})\rangle \; .
\eq
Thus, longitudinal momentum fractions $x_i = k_{i}^{+}/ P^+$ are
Lorentz-boost invariant. A perfectly relativistic state should 
have a dispersion of the form
\be
{\cal M}^2 = 2 P^+ P^- - {\bf P}^2 \label{disper}
\eq
and so diagonalising $\hat{P}^-$ in a basis of fixed $P^+$ and ${\bf P}$ is
equivalent to finding the masses ${\cal M}$ of boundstates. Because
of the transverse lattice however, there will be corrections
to (\ref{disper}) (see later) which we will try to minimize as part
of the optimization of Lorentz covariance.

\section{Heavy Sources and Winding Modes}

Pure QCD is characterised by a single scale in terms of which
all dimensionful quantities, such as hadron masses, can be expressed.
A convenient measure of this scale is the string tension $\sigma$, the
asymptotic slope of the linearly-rising confining potential between heavy
sources. Moreover, by measuring this slope in both transverse
lattice directions ${\bf x}$ and the continuum space direction $x^3$,
and requiring it to be rotationally invariant, we fix the lattice
spacing $a$ in dimensionful units.

\subsection{Winding Modes}

In the transverse directions, a more accurate measurement of
$\sigma$ can be made by examining winding modes rather than
heavy sources. By making the transverse lattice compact in direction
$\hat{\bf \lambda}$ say
\be 
{\bf x} \equiv {\bf x} + a \hat{\bf \lambda} D_{\lambda} \ ,
\eq
where $D_{\lambda}$ is the number of transverse links in direction 
$\lambda$,
we can construct a basis of
Fock states that wind around these directions, e.g. 
\be
         \Tr\left\{ a_{\lambda_1}^\da(k_{1}^{+},{\bf x}) \, 
        a_{\lambda_2}^\da(k_{2}^{+},{\bf x} + a\hat{\bf \lambda}_1)
        \cdots  a_{\lambda_p}^\da(k_{p}^{+},{\bf x}+
                                     a\hat{\bf \lambda}D_{\lambda}-
                                     a\hat{\bf \lambda}_p)\right\} 
        \left|0\right\rangle
        \;   \label{wind}
\eq
has winding number 1 in direction $\hat{\bf \lambda}$.
The mass spectrum of such winding modes should rise linearly with
$D_{\lambda}$
with slope $\sigma$ as $D_{\lambda} \to \infty$. 
Unlike the potential between heavy
sources,
there are no `endpoint' effects, and so the asymptotic linear rise
should set in more quickly, especially for the lowest mass eigenvalue
\be
{\cal M} \to \sigma_{\rm T} | {\bf n} | a \ \ , \ \ |{\bf n}| \to \infty \ ,
\label{wind}
\eq
where ${\bf n} = (D_1 n_1, D_2 n_2)$ and $n_1$, $n_2$ are winding
numbers in directions $(1,0)$ and $(0,1)$. (Note that we cannot
construct winding modes around a compactified $x^3$ as this clashes
with the choice of LF coordinates). The suffix `T' indicates
a transverse measurement of $\sigma$.

\subsection{Heavy Sources}

To measure the potential between two heavy sources in transverse lattice
gauge theory \cite{burk}, we will start with a heavy scalar field 
$\phi(x^+,x^-,{\bf x})$ of
large mass $\rho$ in the fundamental representation of $SU(N)$.
In addition to the pure glue `link-link' interactions, we must now
also consider interactions with $\phi$'s. 
The simplest are
\be
\left(\partial_\alpha \phi^\da
-i \phi^\da A_\alpha \right) \left( \partial_\alpha \phi
+i  A_\alpha \phi \right)
 \ \ , \ \ - \rho^2 \phi^\da \phi
\eq
such that $\phi$ couples to $M$ via $A_{\alpha}$.\footnote{The full
set of operators allowed at leading order of the colour-dielectric
expansion is described in ref.\cite{us5}, but we skip the details
here.}
 In LF
gauge $A_{-} = 0$ this leads to the substitution
\begin{eqnarray}
J^+ & \to & J^{+}_{\rm pure \ glue}[M] + J^{+}_{\rm source} [\phi] \\
(J^{+}_{\rm source})_{ij} & = & -i \partial_{-} \phi_i \phi^{*}_{j}
         +i \phi_i \partial_- \phi^{*}_{j}
\end{eqnarray}

Now for some kinematics:  
Let $P^\alpha_{\rm full}$ represent the full 2-momentum
of a system containing $h$ heavy particles.  
It is convenient to split the full momentum into a ``heavy'' part
plus a ``residual'' part $P^\alpha$ due to interactions,
\be
        P^\alpha_{\rm full} = \rho h v^\alpha + P^\alpha \; , 
        \label{resid}
\eq
%
where $v^\alpha$ is the covariant velocity of the
heavy sources, $v^\alpha v_\alpha =1$. Note that $P^{\alpha}$, as
defined here, is not positive definite.  
The full invariant mass-squared (at ${\bf P} = 0$) is
\be
 {\cal M}^2 = 2 P_{\rm full}^+ P_{\rm full}^- =
      \left(h\rho\right)^2 +2 h \rho v^+ P^- +
        2 P^+\left( P^- + h \rho v^- \right)
\eq
The choice of $v^+$ is arbitrary and it is convenient to choose it
such that $P^+=0$.  Consequently, $v^+ P^-$ is just the shift of the 
full invariant mass $\cal M$ due to the interactions:
\be
  {\cal M} = h \rho +v^+ P^- +O\left(1/\rho\right)\; .
\eq
Thus, $v^+ P^-$ is the usual energy associated with
the heavy quark potential.

\subsection{Fock Space}

We define creation-annihilation operators associated
with the heavy field:
\be
  \phi_i(x^+=0,x^-,{\bf x}) = \frac{1}{\sqrt{4 \pi}}\int_{-\infty}^\infty 
        \frac{dk}{\sqrt{\rho v^+ + k}} \left(
         b_i(k,{\bf x})\,{\rm  e}^{-i (v^+  \rho+ k) x^-}
         + d_{i}^{*}(k, {\bf x})\,{\rm  e}^{i (v^+ 
                 \rho + k) x^-}\right)
           \label{phi} 
\eq
\be
\left[b_{i}(k,{\bf x}), 
        b_{j}^{*}(\tilde{k}, {\bf y})\right] 
         =  \delta_{ij}\, {\bf \delta_{x, y}}\,\delta(k-\tilde{k}) \;,
  \ \ {\rm etc}.
\eq
The ${\rm  e}^{i \rho v_\alpha x^\alpha}$ term removes an 
overall $\rho v^\alpha$ from the 2-momentum.
$b_{i}^{*}(k, {\bf x})$ creates a source particle at site ${\bf x}$
  carrying
residual momentum $k$, while $d^*$ creates an anti-particle. The 
residual gauge invariance in $A_{-} = 0$ gauge again leads to
confinement into singlet states. An example of a gauge-invariant
Fock state with two co-moving sources at ${\bf x}$ and ${\bf y}$
respectively, maintaining a fixed a $x^3$-separation $L$ is
\be
\int_{-\infty}^{\infty} dl {\rm e}^{2{\rm i}Llv^-}
   b^{*}_{i}(l-P_{\rm link}/2,{\bf x})\, a^{*}_{ij}(k_1, {\bf x})\cdots 
a^{*}_{mn}(k_p, {\bf y})\, d^{*}_{n}(-l-P_{\rm link}/2, {\bf y}) |0>
\eq
where $\sum_{i=1}^{p} k_i = P_{\rm link}$. To obtain LF
eigenfunctions of $\hat{P}^-$ we must take linear combinations of
all possible $P_{\rm link}>0$. In practice this requires us to introduce
a high-energy cut-off on $P_{\rm link}$, in addition to the
DLCQ one, $K$. $v^+ \hat{P}^-$ is then a finite dimensional matrix
(at least for the connected Fock states that dominate in the large-$N$
  limit) whose eigenvalues  may be extrapolated to infinite cutoff.

The lowest eigenvalue of $v^+ \hat{P}^-$ can be compared with the
popular phenomenological form of the potential between heavy sources of
spatial separation $R$
\begin{eqnarray}
V(R) & = & \sigma R + c_1 + {c_2 \over R} \label{so} \\
R & = & \sqrt{a^2 |{\bf n}|^2 + L^2} \ \ \ {\rm Source} \ \ {\rm
Separation} ,
\nonumber
\end{eqnarray}
where $a {\bf n} = {\bf y} - {\bf x}$ and $L = y^3 - x^3$.
For generic couplings in $\hat{P}^-$ the eigenvalues will not simply
be  a function of $R$, i.e. $\sigma, c_1, c_2$ extracted from
(\ref{so}) will depend upon $L$ and $|n|$ separately.
Once rotational invariance is restored, 
the value of $\sigma$ can be set from experiment by the masses
and decays of heavy mesons using $V(R)$ in a conventional 
Schrodinger equation. A typical value is $\sqrt{\sigma} \sim 440 MeV$,
though one must remember that this will includes effects of 
light-quarks which are neglected in pure gauge theory. Higher
eigenvalues of $v^+ \hat{P}^-$ should correspond to hybrid heavy
mesons. 

\subsection{Setting the Scales}

We now have everything necessary for performing a first-principles
calculation. Let us now combine results from each sector --- pure
glue,
winding, heavy sources.

\noindent \underline{Pure Glue}

For generic couplings in $\hat{P}^-$, the glueball eigenstates at
fixed momenta $(P^+ , {\bf P})$ have  eigenvalues  which we may
expand in powers of transverse momentum thus
\be
2P^+ P^- = G^2 N \left( \dm^{2}_{0} 
+ \dm_{1}^{2}\, a^2
      |{\bf P}|^2 + 2\overline{\dm}^{2}_{1} a^2 P^1 P^2 +
\dm_{2}^{2}\, a^4 |{\bf P}|^4 + 
             \cdots \right) \label{latshell}\; .
\eq
Here we have chosen to factor out the coupling $G^2$ as an overall
mass scale ($G$ has units of energy). $\dm_0, \dm_1, \overline{\dm}_2,
\cdots$ are then dimensionless numbers which we calculate when
diagonalising $\hat{P}^-$. Eq.(\ref{latshell}) is not, of course,
in the form of a relativistic dispersion relation (\ref{disper}) in general.
The coefficients, however, are functions of the couplings at our
disposal and may be adjusted to regain a relativistic form.

\noindent \underline{Winding Modes}

Let us write the lowest winding eigenvalue (\ref{wind}) as
\be
2P^+ P^- = a^2 \sigma_{T}^{2} |{\bf n}|^2 = G^2 N {\cal W} |{\bf n}|^2
\ \ , \ \ |{\bf n}| \to \infty \label{winder} \ .
\eq
Again, we factor out $G^2$ to set the scale, and ${\cal W}$ is a 
dimensionless measured quantity dependent on the
couplings.

\noindent \underline{Heavy Sources}

Similarly, at fixed sources separation $R$
\be
v^+ P^- = \sigma R = G^2 N {\cal S} R \ \ , \ \ R \to \infty \ .
\label{heavy}
\eq
${\cal S}$ is the dimensionless measured quantity that we can adjust
with the couplings in $\hat{P}^-$.

The following conditions are necessary for Lorentz covariance of
eigenstates
\begin{eqnarray}
\sigma_T & = & \sigma = {\rm constant} \label{ten} \\
\dm_{1}^{2}\, a^2 G^2 N -1 & = & 0 \label{cee}\\
\overline{\dm}_{1} & = & 0 \label{cross}
\end{eqnarray}
Of course, there are further conditions that one could 
impose on solutions. But as we shall see in the next lecture, in
practice 
these seem to be the most significant when trying to remove
lattice discretization errors to a first approximation.

To test for (\ref{cee}) we need the dimensionless combination
$a^2 G^2 N = {\cal W}/{\cal S}^2$ that follows from 
(\ref{winder})(\ref{heavy}). Combining (\ref{latshell})(\ref{heavy}) allows
us to express boundstate masses in units of the string tension
\be
2P^+ P^- ({\bf P}=0) = {\cal M}^2 = {\dm^{2}_{0} \sigma \over {\cal
S}}
\eq
and also the lattice spacing itself $a = \sqrt{{\cal W}/{\cal S}\sigma}$.

\section{Example -- Large-$N$ QCD in $2+1$ Dimensions}

We need to do a test calculation that can be accurately compared with
known results and is relevant to non-abelian gauge theories in 3+1 dimensions. 
$SU(\infty)$ gauge theory in $2+1$ dimensions provides an ideal
test. It is physically very similar to $SU(3)$ pure gauge theory in 
$3+1$ dimensions,
having dynamically-generated 
linear confinement and  a discrete spectrum of glueballs;  
it is even quantitatively similar.
But factorization of colour Traces in the Fock space states in  the 
large-$N$ limit and 
having only one transverse dimension make an accurate
calculation feasible. Moreover, good Euclidean lattice Monte Carlo
(ELMC)
data has recently become available
for this problem \cite{teper}, which we
can take as `the right answer'.\footnote{The calculations of 
Teper are based on the traditional Wilson Euclidean lattice 
path integral formulation of gauge theory. The large $N$ calculations
were in fact performed about the same time
as the light-front work, and to some extent motivated by it.} 
The results presented here have been obtained with B. van de Sande
and are new or improvements on our previously published
work \cite{us3}.

The discussion of the previous lectures is trivially adapted to
$2+1$ dimensions. Here $x^{\pm} = (x^0 \pm x^2)/\sqrt{2}$ etc. and
$x^1$ is the single transverse lattice dimension.
Up order order $M^4$, we have in the large
$N$ limit with $G^2 N$ finite (suppressing
transverse indices for clarity)
\begin{eqnarray}
 P^-  & = & \int dx^- 
         - {G^2 \over 4} \Tr\left\{  J^{+}  
        \frac{1}{\partial_{-}^{2}} J^{+} \right\}
         + {G^2 \over 4N} \Tr  J^{+} \frac{1}{\partial_{-}^{2}}
          \Tr J^{+} 
        + \mu^2  \Tr\left\{MM^{\da}\right\} \nonumber \\ 
        && + {\lambda_1 \over a N}
        \Tr\left\{ M M^{\da}MM^{\da} \right\} + {\lambda_2 \over a N}  
        \Tr\left\{ MMM^{\da} M^{\da} \right\} +  {\lambda_3 \over a N^2} 
        \left( \Tr\left\{ MM^{\da} \right\} \right)^2 
\end{eqnarray}
where, in anticipation of always taking the ${\cal L} \to \infty$
limit,  we have also used power-counting in longitudinal coordinates
to limit the number of terms. There are in principle an infinite
number of different products of individually gauge-invariant
operators (like the $\lambda_3$ term), obtained by separating the
operators in the transverse direction. We have also assumed
a degree of transverse locality by keeping only the most local
product to a first approximation. This approximation, like the
colour-dielectric expansion in gauge-invariant powers of
$M$, can of course be systematically investigated.

To maintain a trivial LF vacuum, we
shall be considering the region $\mu^2 > 0$ with tachyon-free physical
spectrum. $G$ has dimensions of energy and will be used to express 
dimensionful quantities;
it is convenient to form dimensionless versions of all the other
couplings
\be
        m^2 = {\mu^2 \over G^2 N}  \; , \;\;\;\; 
        \newl_i = {\lambda_i \over a G^2 N}  \; .
\eq
The basic technique we follow is to search the space $\{m,\newl_1,
\newl_2,\newl_3 \}$ for a one parameter trajectory on which
observables show
enhancement of Lorentz covariance --- a `Lorentz trajectory'. Moving
along this trajectory should correspond to changing the spacing
$a$, eventually taking us to the continuum limit. We will, in fact,
be prevented from reaching $a=0$ by our self-imposed restriction
$m^2 > 0$. Nevertheless, assuming gauge and Lorentz covariance
{\em define} pure QCD, observables should be invariant along the (exact)
Lorentz trajectory.\footnote{The idea is similar to the renormalised
trajectory in the renormalisation group associated to a fixed point
with one relevant direction. However we are not performing any
explicit RG transformations here!} Of course, we can only obtain
some approximation to such a trajectory with a finite number of
couplings.

\begin{figure}
\centering
$\chi^{2}_{\rm min}$
\BoxedEPSF{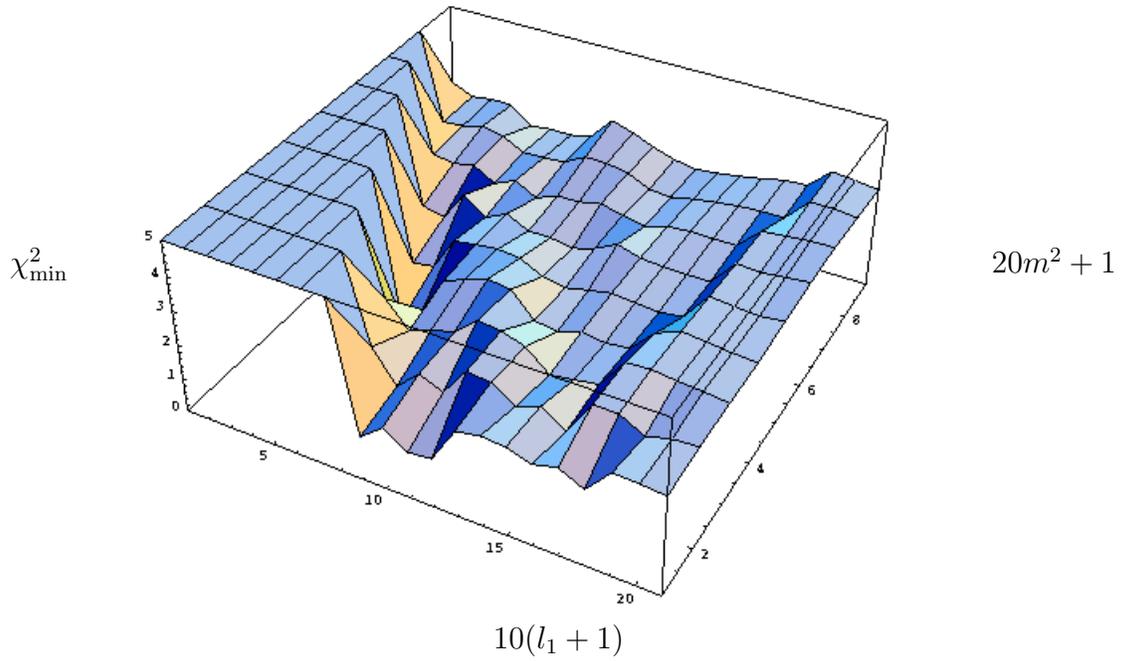 scaled 600}
$20m^2+1$\\
\hspace{-0.1in} $10(\newl_1+1)$ \\
\vspace{5mm} 
\caption{$\chi^2$-chart for $\newl_1$. 
\label{fig2}}
\end{figure}

\begin{figure}
\centering
$20m^2 +1$.
\BoxedEPSF{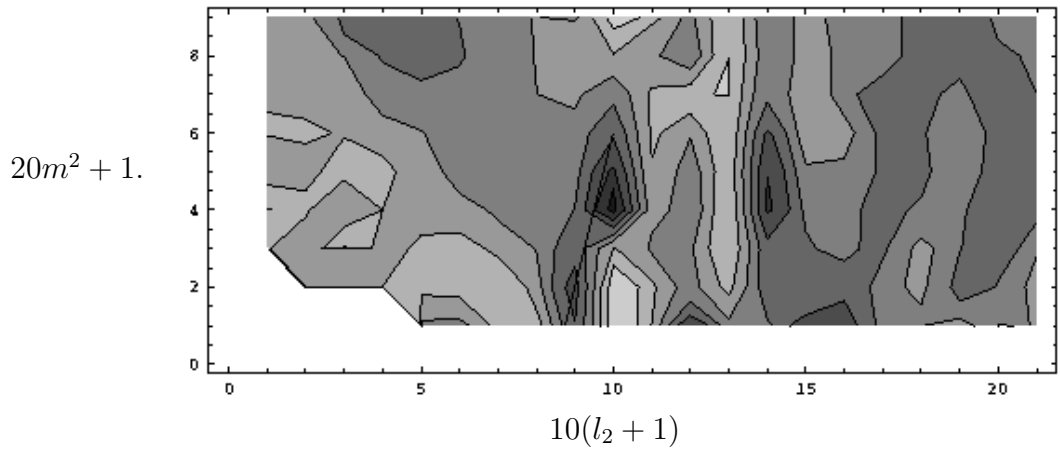 scaled 700}
\\ \hspace{0.5in}$10(\newl_2 + 1)$ \\
\vspace{5mm} 
\caption{$\chi^2$-chart for $\newl_2$. Darker shades
correspond
to lower $\chi^{2}_{\min}$.
\label{fig2}}
\end{figure}

To get some intuition for the space of couplings, let us construct 
charts that display the `size' of violations of Lorentz covariance. We
can use a $\chi^2$-test based on the variables
(\ref{ten}))(\ref{cee}) (condition (\ref{cross}) is absent in
$2+1$-dimensions). Condition (\ref{cee}) in fact provides a number of
variables, one for each boundstate (glueball) in the low-lying
spectrum.
To begin with, let us set the lowest glueball mass to the known
value ${\cal M} = 4.05 \sqrt{\sigma}$ \cite{teper}, and try to get an isotropic
speed of light (\ref{cee}) for this and the heavier low-lying
glueballs. With a particular choice of variance assignments in the
$\chi^2$-test, we can make plots of $\{ \chi^{2}_{\rm min}, m^2,
\newl_i \}$ for each $\newl_i$, where $\chi^{2}_{\rm min}$ is the
minimum value of $\chi^2$ with respect to the couplings except
$m^2$ and $l_i$. For technical reasons $\newl_3$ almost completely
decouples (we can set it to almost any fixed large value, see
ref.\cite{us3})
so we only display charts for $l_1$ (fig.1) and $l_2$ (fig.2).

They each show an unique, narrow valley, running from small to large
$m^2$, at the bottom of which Lorentz covariance is
optimised. One finds that changing
the details of the $\chi^2$ test changes these charts somewhat, {\em except}
in the neighborhood of the valley in each case. Thus one finds a
degree of universality in the results. If there were no Lorentz
trajectory, one would not expect to obtain a robust valley.
Having roughly located the candidate Lorentz trajectory, one can
increase the resolution in its neighborhood, and perform more
efficient iterative searches for the bottom of the valley.

\begin{figure}
\centering
${a \sqrt{\sigma} }$
\BoxedEPSF{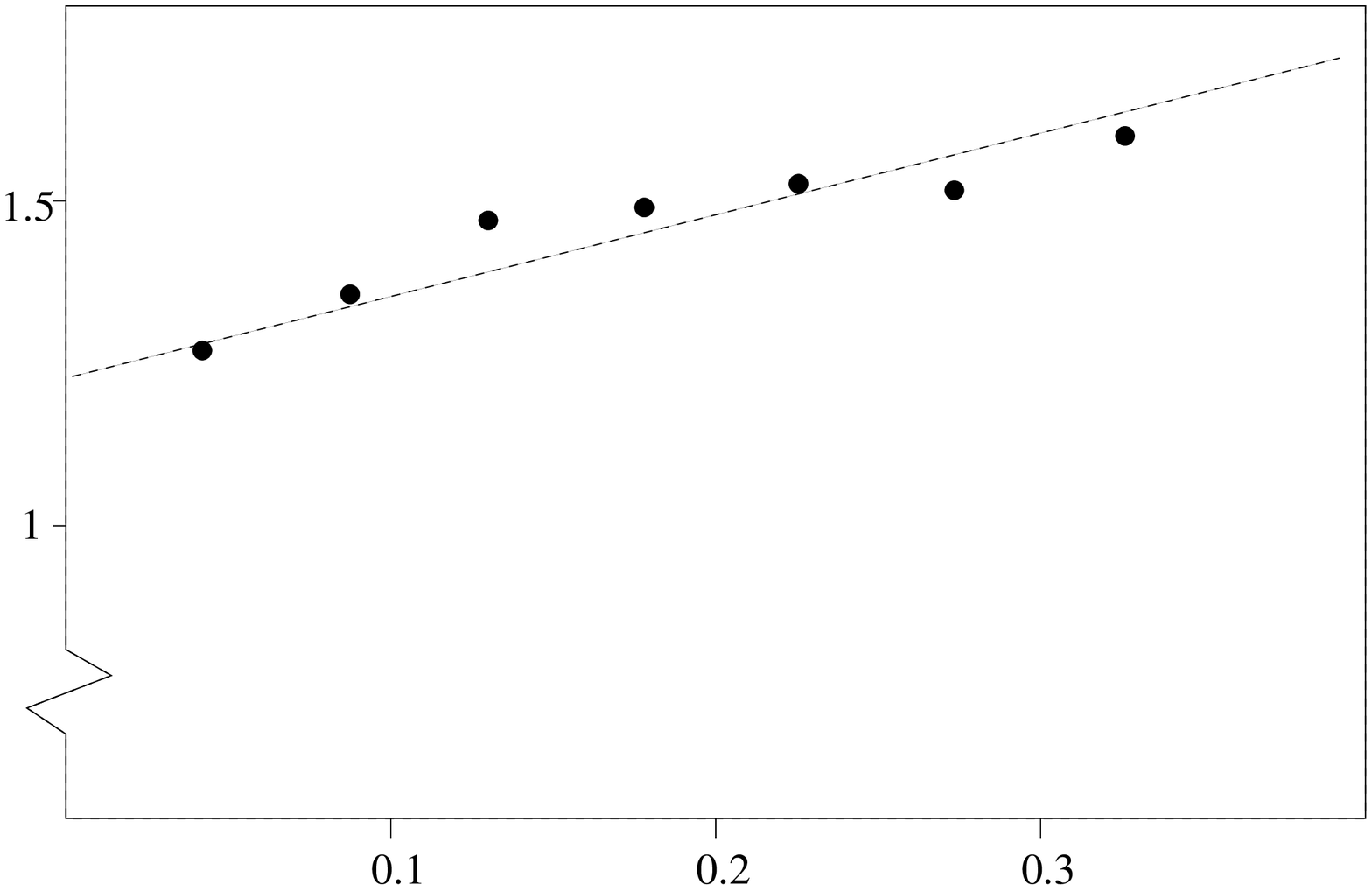 scaled 500}
\\ \hspace{0.5in}$m$
\caption{
Variation of the transverse lattice spacing along the Lorentz
trajectory. The fit is $1.275 m + 1.23$. 
\label{fig2}}
\end{figure}

\begin{figure}
\centering
${{\cal M} \over \sqrt{\sigma}}$
\BoxedEPSF{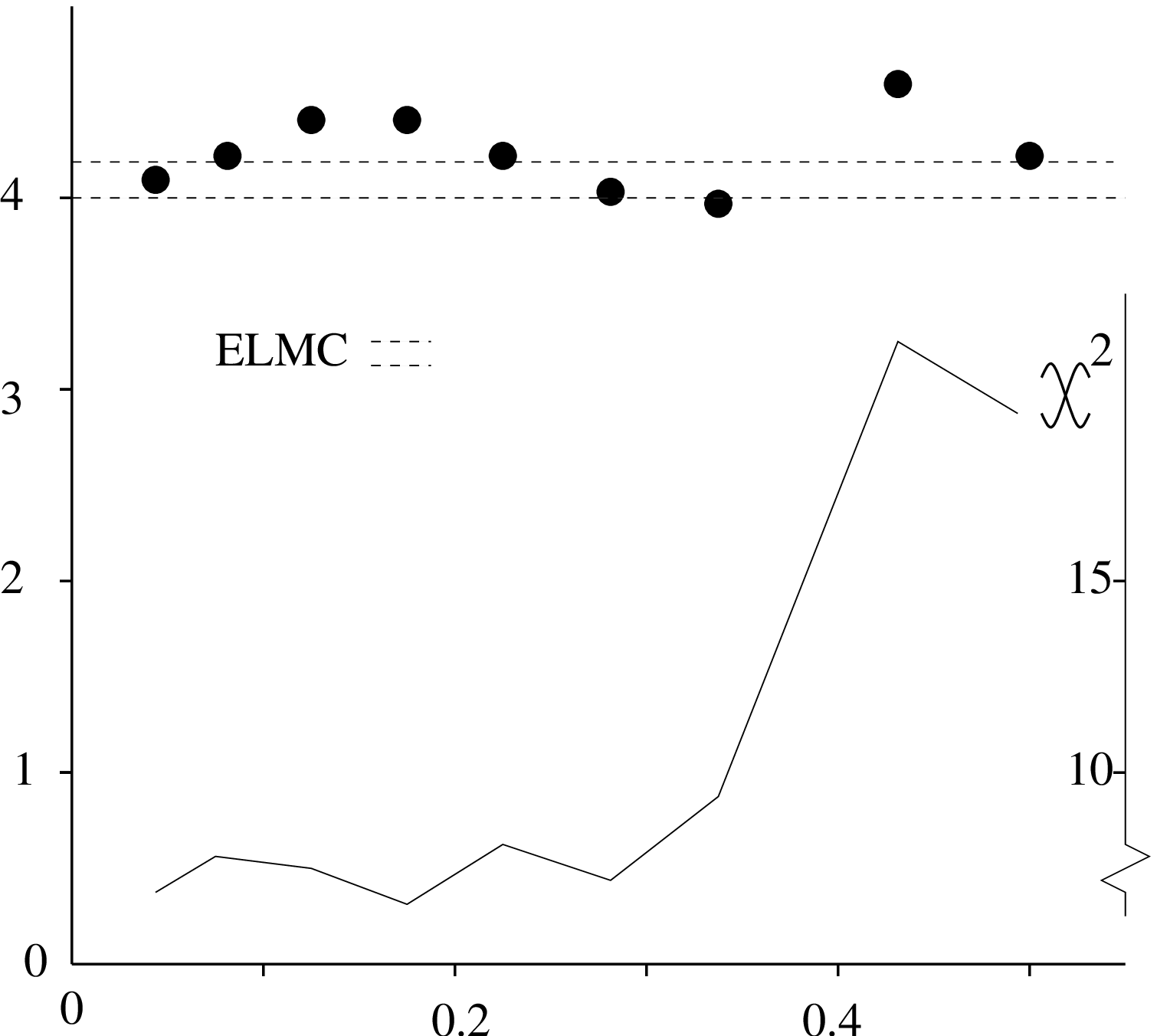 scaled 700}
\\ \hspace{0.5in}$m$
\caption{The variation of the lightest glueball mass along the
Lorentz trajectory (together with the associated variation of the
$\chi^2$).
\label{fig6}}
\end{figure}

In a first-principles calculation we cannot use the result for the
lightest glueball mass in $\sigma$ units --- this should be predicted
by the method. Performing a new $\chi^2$ search in the neighborhood of
the valley, but now including a variable to test (\ref{ten}), 
one finds a similar position for the valley bottom, without having to
input the lightest glueball mass. The $\chi^2$ per degree of
freedom is always about 1 near the Lorentz trajectory, as it should be
for reasonable criteria.
Results as one moves along the
valley bottom are displayed in fig.3 and fig.4. 
We see that the lattice spacing
gradually decreases with $m^2$ as one moves along the Lorentz
trajectory,
but never becomes zero for $m^2 >0$. Despite fluctuations, the
lightest
glueball mass scales in a way roughly consistent with the correct
continuum answer.

\begin{figure}
\centering
$\frac{\cal M}{{\cal M}_{O^{++}}}$
\BoxedEPSF{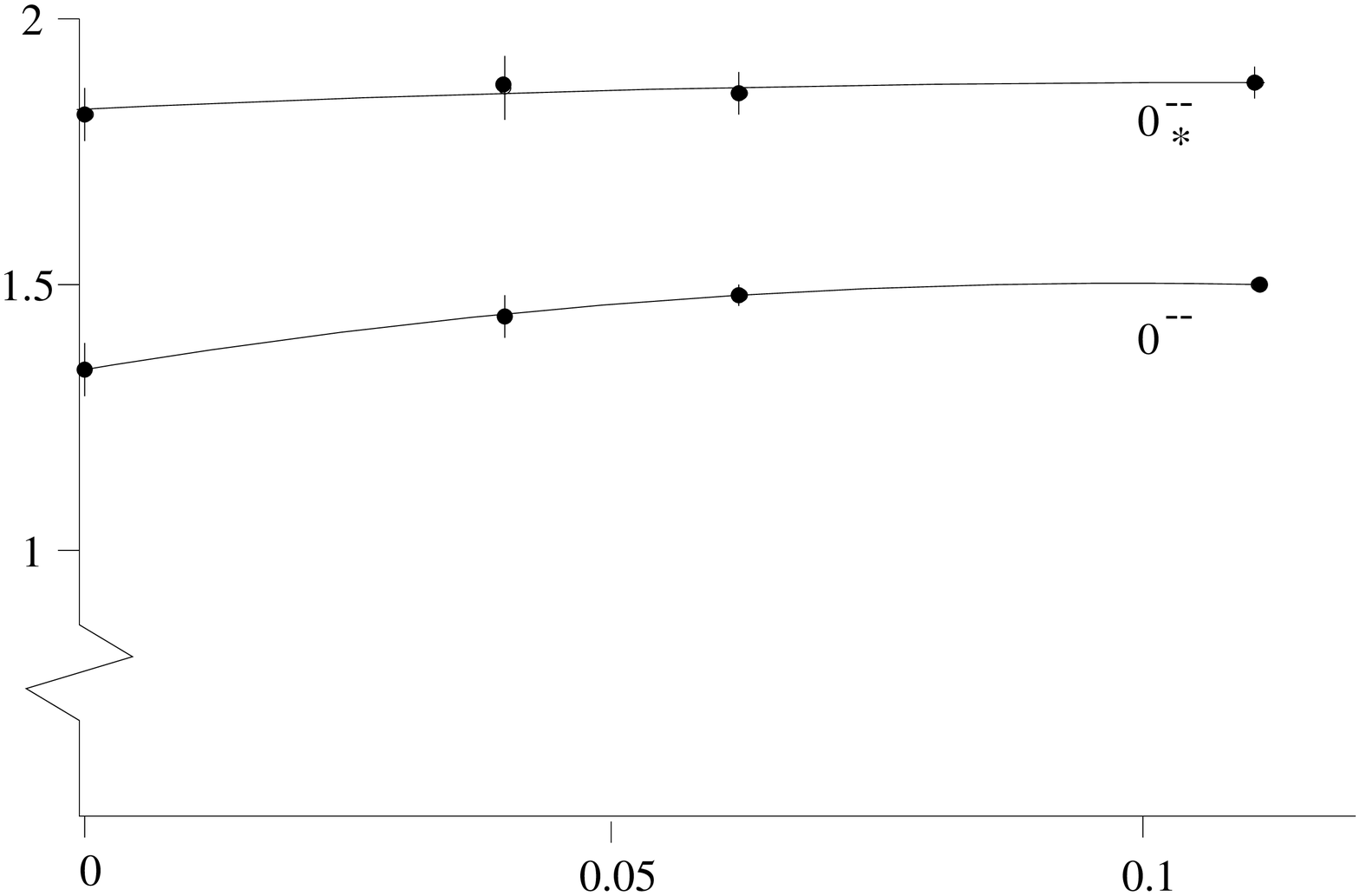 scaled 400}
\\ \hspace{0.5in}$1/N^{2}$
\caption{Variation (or lack of it!) 
of the glueball mass ratios $0^{--}/0^{++}$ and
$0^{--}_{*}/0^{++}$ with $N$. The fits are $1.349 + 2.914/N^{2}
-13.968/N^{4}$ for $0^{--}$ and $1.824 + 1.274/N^{2}
-7.143/N^{4}$ for $0_{*}^{--}$. The error on the $N=\infty$
results is
from extrapolation to ${\cal L} = \infty$.
\label{fig2}}
\end{figure}

Although the lightest glueball 
(a $J^{PC} =0^{++}$) is behaving covariantly along the Lorentz
trajectory --- the speed of light deduced from the $0^{++}$'s dispersion 
is isotropic to with \% 2-3 
--- fluctuations are still present in fig.4.
These are mostly due to the difficulty
in accurately establishing the scale $\sigma$. Most of the
fluctuations tend to cancel if we consider ratios of glueball
masses, and good scaling is obtained along the
Lorentz trajectory \cite{us3}. For higher glueballs
 with nearly-covariant
wavefunctions, such as the $0^{--}$ and its excited state
$0^{--}_{*}$,
this provides rather accurate mass ratio determinations in the
large-$N$ limit; see
figure 5. Finite-$N$ results from
conventional ELMC in ref.\cite{teper} were
fit to $A + B/N^{2}$, at
confidence levels of order $\% 50-70$. Including our $N = \infty$
data, and fitting to $A + B/N^{2} + C/N^{4}$, improves this 
to
$\% 95$ for $0^{--}_{*}/0^{++}$ and $> \% 99$ for $0^{--}/0^{++}$
glueball
mass ratios!

\begin{figure}
\centering
${v^+ P^- \over \sqrt{G^2 N}}$
\BoxedEPSF{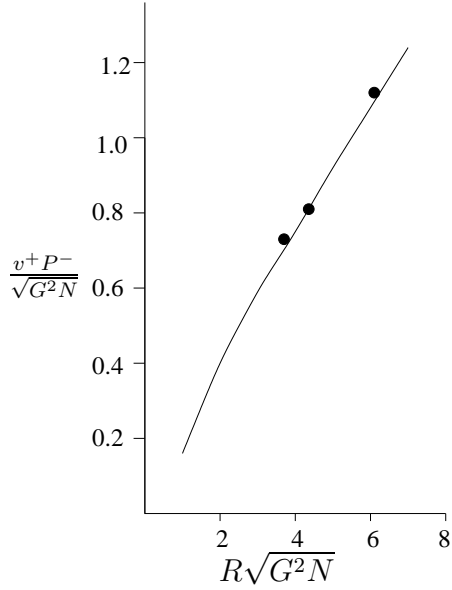 scaled 500}
\\ \hspace{0.5in}$R \sqrt{G^2 N}$
\caption{
The heavy-source potential. Solid line is fit to potential for
sources with $x^2$-separation only; data points are values at one-link
transverse separation and $x^2$-separation $L \sqrt{G^2 N} = 0, 2.5, 5$.
\label{fig2}}
\end{figure}

The heavy-source potential calculated at a point on the Lorentz 
trajectory is displayed in fig.6. It shows good
restoration of spatial symmetry. The potential in the 
continuum spatial direction $x^2$ is a fit to 
\be
v^+ P^-   = 0.154 L G^2 N + 0.183 \sqrt{G^2 N} - {0.178 \over L} \label{care}
\eq
at the point on the Lorentz trajectory of lowest overall $\chi^2$.
One must be careful when interpreting (\ref{care}) 
since the Coulomb potential in $2+1$
dimensions
is logarithmic. The form (\ref{care}) 
should be appropriate except at the very smallest
$L$, where Coulomb corrections are expected. The $1/L$ term is
a universal correction expected on the grounds of models of flux-string
oscillations \cite{flux}. Universality implies that its coefficient  
should be invariant along the Lorentz trajectory. In
reality, we find that it drifts slowly,  a symptom that our approximation
to the Lorentz trajectory is not an exact scaling trajectory and/or
the form (\ref{care}) is not sufficient to fit the potential.
The coefficient 0.178 at the lowest $\chi^2$ 
is nevertheless close to the theoretical
value $d\pi /24$, where $d$ is the number of transverse dimensions
in which an ideal thin-string can oscillate ($d=1$ here). 
But it  is larger  by $40 \%$  (0.178 gives $d \sim 1.4$).
We found the same excess by another method to measure the
`central charge' $d+2$, using the density of glueball states \cite{us2}.
In ref.\cite{us2} we observed that this excess in the central charge
comes from longitudinal degrees of freedom in the large-$N$ 
QCD flux string, a mode
of oscillation which an ideal thin-string does not possess.

One may wonder whether results really do improve as more operators
are added to the hamiltonian.
For a given $\chi^2$ test, adding more operators to the hamiltonian
will inevitably bring one closer to the Lorentz trajectory as measured
by this test. 
We see from fig. 7 that the masses of glueballs,
which are not part of the test, rapidly approach the
correct values as more operators are added and covariance is 
improved.\footnote{The
$2^{-+}$ state appears to overshoot the correct mass. It is the
only state whose wavefunction actually becomes less covariant when $\lambda_1$ 
is turned on.}

\begin{figure}
\centering
\BoxedEPSF{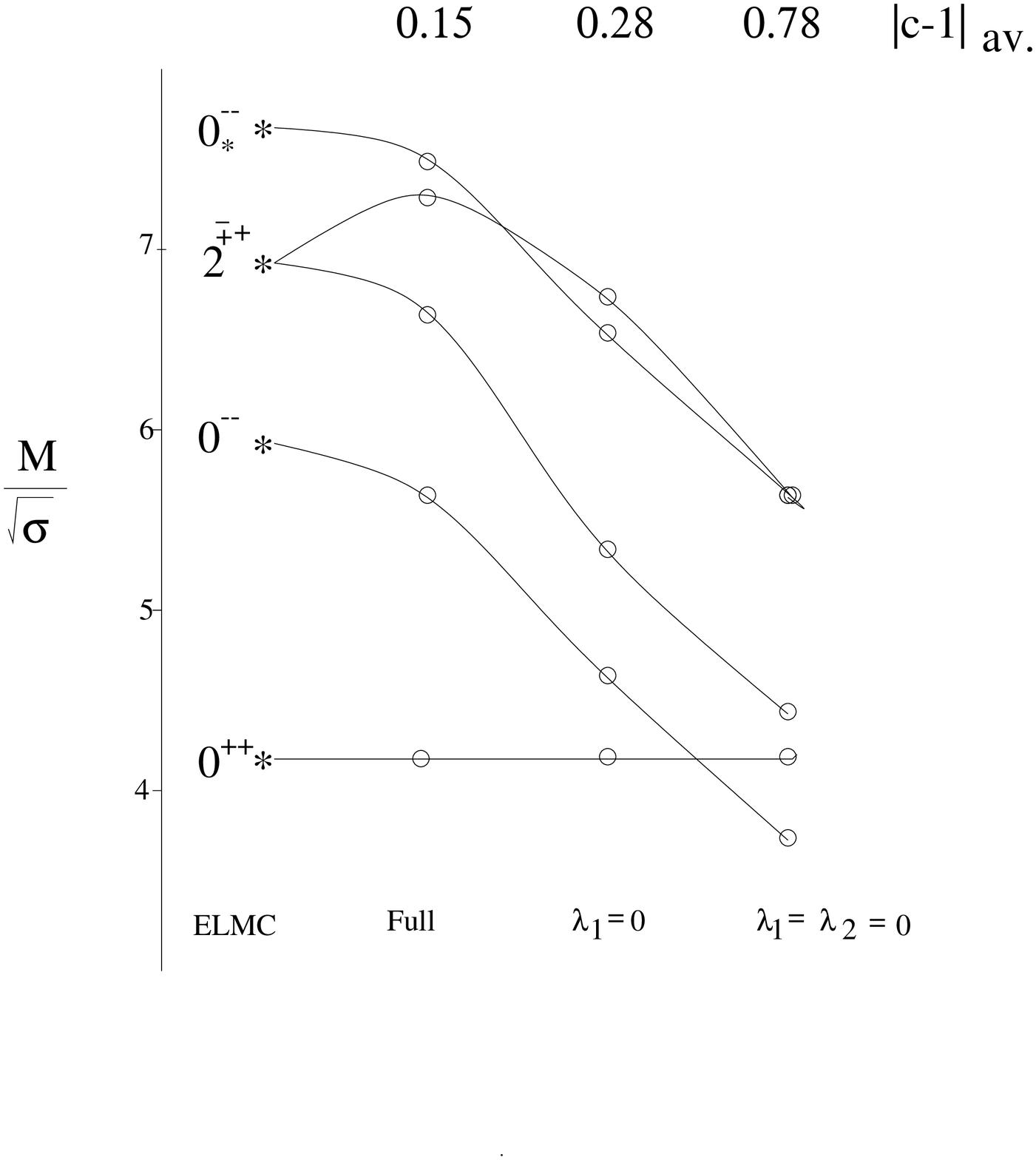 scaled 400}
\caption{Variation of glueball mass ratios to the lightest mass ($0^{++}$), as
more operators are included in the hamiltonian. $|c-1|_{\rm av}$ 
measures the average deviation from 1 of the speed of light in the
transverse direction, for the displayed glueballs.
\label{fig4}}  
\end{figure}

\subsection{Future Work}

The transverse lattice idea is already quite old \cite{bard1}.
The breakthrough, that has been
made over the last year or so, was to
show that in $2+1$ and $3+1$ dimensions large-$N$ gauge theory 
exhibit a unique Lorentz covariant scaling trajectory on coarse
transverse lattices \cite{us3,us4,us5}. Glueball masses on this trajectory are 
consistent with known results. 
Thus there is good reason to 
believe that the Lorentz-invariant
wavefunctions obtained with this method are accurate also. 
They are completely new results, essentially unobtainable within any other
quantisation scheme. It will be interesting to see what implications
they have for experiment. The next step towards this goal is to
couple these pure gauge theories to propagating quarks
\cite{bard1,burk2}, 
and then
enforce the Lorentz and chiral symmetries that define QCD to determine
the couplings of the hamiltonian. We also need to understand better
how to extrapolate results into the small $a$ region. This is
important
not only formally for the existence of the continuum limit, but to
compare calculations of hadronic structure with hard process
experiments
where the usual factorization theorems simplify the analysis.
We are only at the beginning of the development of this subject,
and it is difficult to know how far it will progress, but there does 
appear to be too much truth in it to ignore.

\vspace{5mm}

{\bf Acknowledments} I would like to thank Prof. Ji and Prof. Min
for inviting me to give these lectures and for their hospitality in
Korea. This work was supported by PPARC grant 
No.\ GR/LO3965.

\vfil
\end{document}